%% file: main.tex
\documentclass[aps,jcp,preprint,superscriptaddress,floatfix]{revtex4-1}
\usepackage{epsfig}
\usepackage{graphicx}
\usepackage{color}
\usepackage{dcolumn} 
\usepackage{bm} 
\usepackage{multirow} 
\usepackage{pifont} 
\usepackage{amsmath} 
\usepackage{subfigure}
\usepackage{float}
\usepackage{bm}
\usepackage{modiagram}
\newcommand{\RuBpy}{[Ru$^{\rm II}$(Bpy)$_3$]$^{2+}$}
\newcommand{\RuAm}{[Ru$^{\rm III}$(NH$_3$)$_6$]$^{3+}$}
\newcommand{\RuCN}{[Ru$^{\rm II}$(CN)$_6$]$^{4-}$}

\begin{document}

\author{Daniel R. Nascimento}
\email{daniel.nascimentodasilva@pnnl.gov}
\affiliation{
         Physical and Computational Sciences Directorate, 
         Pacific Northwest National Laboratory,
         Richland, Washington 99352, USA.}
\author{Elisa Biasin}
\affiliation{Stanford PULSE Institute, SLAC National Accelerator Laboratory, Menlo Park, California 94025, USA.}
\author{Benjamin Poulter}
\affiliation{Department of Chemistry, University of
Washington, Seattle, Washington 98195, USA.}
\author{Munira Khalil}
\affiliation{Department of Chemistry, University of
Washington, Seattle, Washington 98195, USA.}
\author{Dimosthenis Sokaras}
\affiliation{Stanford PULSE Institute, SLAC National Accelerator Laboratory, Menlo Park, California 94025, USA.}
\author{Niranjan Govind}
\email{niri.govind@pnnl.gov}
\affiliation{
          Physical and Computational Sciences Directorate, 
          Pacific Northwest National Laboratory,
          Richland, Washington 99352, USA.}

\title{A time-dependent density functional theory protocol for resonant inelastic X-ray scattering calculations}

\begin{abstract}
 We present a time-dependent density functional theory (TDDFT) based approach to compute the light-matter couplings between two different manifolds of excited states relative to a common ground state. These quantities are the necessary ingredients to solve the Kramers--Heisenberg equation for resonant inelastic X-ray scattering (RIXS) and several other types of two-photon spectroscopies. The procedure is based on the pseudo-wavefunction approach, where TDDFT eigenstates are treated as a configuration interaction wavefunction with single excitations, and on the restricted energy window approach, where a manifold of excited states can  be rigorously defined based on the energies of the occupied molecular orbitals involved in the excitation process. We illustrate the applicability of the method by calculating the 2p4d RIXS maps of three representative Ruthenium complexes and comparing them to experimental results. The method is able to accurately capture all the experimental features in all three complexes, with relative energies correct to within 0.6 eV at the cost of two independent TDDFT calculations. 
\end{abstract}

\maketitle

Resonant inelastic X-ray scattering (RIXS) \cite{Gelmukhanov:1999:87,deGroot:2001:1779,Ament:2011:705} is a two-photon scattering process in which a system is excited at a given X-ray absorption region followed by the emission of a lower energy photon, leaving the system at an excited state. Thus, RIXS provides valuable information on the electronic structure of both occupied and virtual states that are not easily accessible by one-photon spectroscopies due to selection rule restrictions.

Although RIXS has been a well established technique in the study of condensed matter systems, as core-excited states in these systems are relatively easy to access without the need of high-intensity light sources \cite{Kotani:2001:203} (due to the high concentration of atomic centers), the continuous advance in light source technologies are allowing RIXS to also flourish in the context of gas/solution-phase molecular spectroscopy \cite{Hennies:2010:193002,Nordgren:2013:0368,Kunnus:2013:16512,Pietzsch:2015:088302,Eckert:2017:6088,Ross:2018:5075,Hahn:2018:9515,Temperton:2019:074701,Fouda:2020:null}. As RIXS enters the domain of larger and more complex molecular systems, the need for reliable and inexpensive electronic structure computational methods to aid in the prediction and interpretation of complicated spectral features becomes crucial.  

Recent years have witnessed the development of several theoretical approaches for RIXS relevant to molecular systems based on wavefunction theories such as damped response and equation-of-motion coupled-cluster \cite{Faber:2019:520,Faber:2020:2642,Nanda:2020:2629,Nanda:2020:244118}, algebraic diagrammatic construction \cite{Rehn:2017:5552}, multiconfigurational SCF
\cite{Josefsson:2012:3565}, and configuration interaction \cite{Maganas:2017:11819} theories. For a broader overview of electronic structure methods in the context of X-ray spectroscopies, we refer the reader to the comprehensive review by Norman and Dreuw \cite{Norman:2018:7208}. Despite their high accuracy and black-box nature, wavefunction-based approaches remain too expensive to be routinely applied to large molecules (more than a few non-hydrogen atoms). Hence, novel approaches based on density functional theory (DFT) are highly desirable, as they are considerably cheaper than their wavefunction counterparts and still provide reasonably good accuracy. 

In the context of DFT, two main approaches to the calculation of RIXS have been employed. The first is built on an independent-particle picture where Kohn--Sham orbitals and their respective energies are used to construct the required transition moments and excitation energies \cite{Hanson-Heine:2017:094106,Fouda:2018:2586,Besley:2020:1306}. While this approach has been applied to several simple systems showing surprisingly accurate results \cite{Hanson-Heine:2017:094106}, an independent-particle model is expected to break down as the level of correlation of the relevant excited states increases. The second approach is to consider a post-DFT single-reference core-excited intermediate state constructed by swapping the occupations of specific core and virtual orbitals. This reference state is then fed into a linear-response time-dependent DFT (TDDFT) procedure, which naturally yields the emission spectra resulting from the re-population of the core hole \cite{Zhang:2012:194306,Zhang:2015:5804}. Although this approach does include excited-state correlations for the final states, the single hand-picked intermediate is not an obvious choice, and is usually insufficient to describe the relevant physics, especially if there is the possibility of many atomic centers contributing to the manifold of intermediate states.

In this Letter, we present a TDDFT protocol to compute the manifold of intermediate and final states (each comprised of a few to several hundred excited states) involved in the RIXS process and their respective light-matter couplings. The procedure is based on the pseudo-wavefunction approach previously employed in the context of derivative couplings calculations \cite{Ou:2015:064114,Ou:2015:7150,Zhang:2015:064109,Alguire:2015:7140}. Since the relevant states are all obtained from a TDDFT procedure, excited-state correlations are captured, and because all intermediate states within an energy window are considered, the present protocol does not suffer from the single intermediate problem. Furthermore, the computational cost is virtually the same as two separate TDDFT calculations. We illustrate the applicability of the method by calculating the 2p4d RIXS maps of three representative ruthenium complexes.

\begin{figure}[!htpb]
\caption{Schematic description of the RIXS process.}
\includegraphics[scale=1.0]{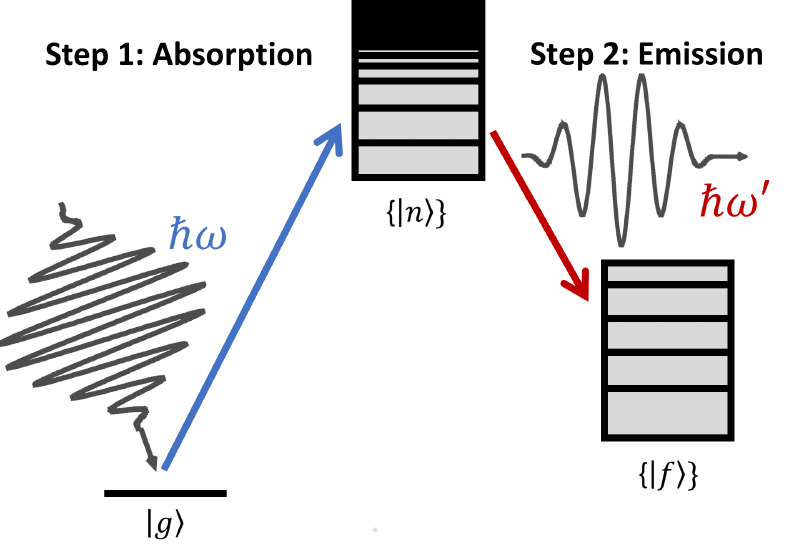} 
\label{FIG:lambda_type}
\end{figure}

The RIXS process can be conceptualized as an infinite set of $\Lambda$-type 3-level systems as illustrated in Figure \ref{FIG:lambda_type}.
The system is initially in its ground state $|g\rangle$, and is subsequently excited into a short-lived intermediate state $|n\rangle$ through absorption of a resonant X-ray photon of energy $\omega$. The system then decays to a final state $|f\rangle$ emitting a photon of energy $\omega'$. If the emitted photon has the same energy as the one absorbed, the final state will be the ground state. In which case, the scattering process is said to be elastic (Rayleigh scattering), and we gain no extra information than what we would have gained from an absorption process. On the other hand, if the process is inelastic, the energy transferred in the process will contain valuable information about the correlations between the two different manifolds of excited states. Thus, the RIXS process is closely related to X-ray Raman Scattering (XRS) \cite{Nordgren:2013:0368}, with the latter referring to the case where the energy of the incident photon does not resonate with any of the intermediate states.  
In the non-relativistic limit, the 2-dimensional RIXS map with respect to the near-resonant absorption frequency $\omega$ and the emission frequency $\omega'$ is given (in atomic units) by the generalized Kramers--Heisenberg (K-H) equation \cite{Kramers:1925:681,Gel'mukhanov:1994:4378,Gelmukhanov:1999:87}

\begin{equation}
    \label{EQN:KHdef}
    \sigma_\theta (\omega',\omega) = \frac{\omega'}{\omega} \sum_{fn} |F_{fn}(\theta)| \left [ \frac{(\omega_n \omega_f \alpha)^2}{(\omega - \omega_n)^2 + \Gamma_n^2/4} \right ]  \times \delta(- \omega_f + \omega - \omega'),
\end{equation}
where $\omega_k$ represents the excitation energy from the ground state $|g\rangle$ to state $|k\rangle$, $\Gamma_k$ is the lifetime broadening associated with $|k\rangle$ that accounts for non-radiative processes not included explicitly in the Hamiltonian, and $F_{fn}(\theta)$ is the polarization-dependent transition amplitude, defined as \cite{Gel'mukhanov:1994:4378}
\begin{equation}
    \label{EQN:rixs_ang}
    F_{fn}(\theta) = \frac{1}{15} \sum_{\xi \xi'} \left [ \left (2 - \frac{1}{2}\sin^2 \theta \right) \left (S_{fn}^{\xi \xi'} \right )^2 + \left( \frac{3}{4} \sin^2 \theta - \frac{1}{2} \right ) \left( S_{fn}^{\xi \xi} S_{fn}^{\xi' \xi'} + S_{fn}^{\xi \xi'} S_{fn}^{\xi' \xi} \right )\right ],
\end{equation}
with
\begin{equation}
    S_{fn}^{\xi \xi'} =  \langle f|\hat{\mathcal{T}}_{\xi}^\dagger|n \rangle \langle n|\hat{\mathcal{T}}_{\xi'}|g \rangle.
\end{equation}
Here, $\theta$ is the angle between the polarization and the propagation direction of the outgoing photon and $\hat{\mathcal{T}}_\xi$ is the $\xi$-component of the coupling operator connecting the photon and electric fields. In order to minimize the effects of elastic scattering, experiments are usually conducted at $\theta$=0.

In eq. \ref{EQN:KHdef}, it is possible to recognize two resonance conditions, the first one is in the denominator which is singular (in the limit $\Gamma_n \to 0$) when the incident energy matches the excitation energy of one of the intermediate states $|n\rangle$, and the second is the Dirac delta function which is unity when the energy transferred in the inelastic process, $\Omega = \omega - \omega'$, matches the excitation energy of one of the many possible final states and vanishes otherwise. Similarly, the numerator of eq. \ref{EQN:KHdef} contains a product of transition moments connecting the ground state to a manifold of intermediate states, which in turn, is coupled to a manifold of final states. Thus, the ingredients required to compute RIXS maps via the K-H equation are the energies and wavefunctions of the ground state, and of the manifold of possible intermediate and final states. Such a computation may in principle seem cumbersome, nonetheless, if we let the relevant excited states be approximated as linear combinations of singly-substituted Kohn--Sham (or Hartree--Fock) determinants where only excitations out of a predefined orbital set, $I$, are allowed, that is,
\begin{equation}
    \label{EQN:cis}
    |k\rangle \approx \sum_{ia} X^k_{ai} \hat{a}_a^\dagger \hat{a}_i |\Phi_{\rm KS}\rangle, \forall i \in I, 
\end{equation}
the creation of several manifolds of excited states can be achieved without too much effort. For instance, the set of occupied orbitals with 2p character in a transition metal may be chosen to generate a manifold of L$_3$-excited states, whereas the set of occupied 1s orbitals will generate a manifold of K-excited states. This constraint can be rigorously achieved by the restricted excitation window approach\cite{Stener:2003:115,Lopata:2012:3284}, where the set of occupied orbitals used in eq. \ref{EQN:cis} is defined by those orbitals within a user-defined energy window, $I = \{i ~|\epsilon_{\rm min} \le \epsilon_i \le \epsilon_{\rm max}\}$ and no restrictions on the target unoccupied states. The states within a given excitation manifold can, then, be obtained by independent TDDFT (or TDHF/CIS) calculations provided that they share the same ground state reference.

After obtaining the excited-states pseudo-wavefunctions, ${\mathbf X}^k$, the couplings needed to solve eq. \ref{EQN:KHdef} may be computed as 
\begin{equation}
    \langle n|\hat{\mathcal T}_\xi|g\rangle = \sum_{ia} T^{\xi}_{ia} X_{ai}^n 
\end{equation}
and
\begin{equation}
    \label{EQN:excoupl}
    \langle f|\hat{\mathcal T}^{\dagger}_\xi|n\rangle = \sum_{ia} X_{ia}^f \left ( \sum_b \tilde{T}^{\xi}_{ab} X_{bi}^n - \sum_j X_{aj}^n \tilde{T}^{\xi}_{ji} \right ),
\end{equation}
where $T^{\xi}_{pq}$ ($\tilde{T}^\xi_{pq}$) is a matrix element of the operator $\hat{\mathcal T}_\xi$ ($\hat{\mathcal T}^{\dagger}_\xi$) in the molecular orbital basis. In the electric dipole approximation, this operator is simply the electric dipole operator expressed, in the length gauge, as $\hat{\mu}_{\xi} = \hat{\mu}_{\xi}^\dagger = -\sum\limits_{pq} \langle \phi_p|{\mathbf r}_\xi|\phi_q\rangle \hat{a}_p^\dagger \hat{a}_q$. It is worth noting that the electric dipole approximation is not always valid within the X-ray regime, thus, the inclusion of higher-order multipoles in the definition of $\hat{\mathcal T}_\xi$ may be necessary.

Note also that eq. \ref{EQN:excoupl} can be, equivalently, recast in terms of excited-state transition density matrices
\begin{eqnarray}
    \gamma_{ba}^{nf} = \sum_{i} X_{bi}^n X_{ia}^f \\
    \gamma_{ij}^{fn} = \sum_{a} X_{ia}^f X_{aj}^n
\end{eqnarray}
as
\begin{equation}
    \label{EQN:excoupl_tdm}
    \langle f|\hat{\mathcal T}_\xi|n\rangle = \sum_{ab} \gamma_{ba}^{nf} T^{\xi}_{ab} - \sum_{ij} \gamma_{ij}^{fn} T^{\xi}_{ji}.
\end{equation}

The protocol proposed herein was implemented in a development version of the {\small NWChem} electronic structure package \cite{Valiev:2010:1477,Apra:2012:184102}, and its validity has been assessed by computing the 2p4d RIXS spectra of three octahedral ruthenium model complexes, namely, tris(2,2'-bipyridyl)ruthenium(II) (\RuBpy), hexacyanoruthenate(II) (\RuCN), and hexaammineruthenium(III) (\RuAm).

\begin{figure}[!htpb]
\caption{Complexes studied in the present work, simplified molecular orbital diagram for a generic octahedral system, and calculated Ru L$_3$-edge absorption spectra for all three complexes. A uniform shift of -2.3 eV and an artificial Lorentzian broadening of 1.5 eV have been applied to the spectra.}
\includegraphics[scale=1.0]{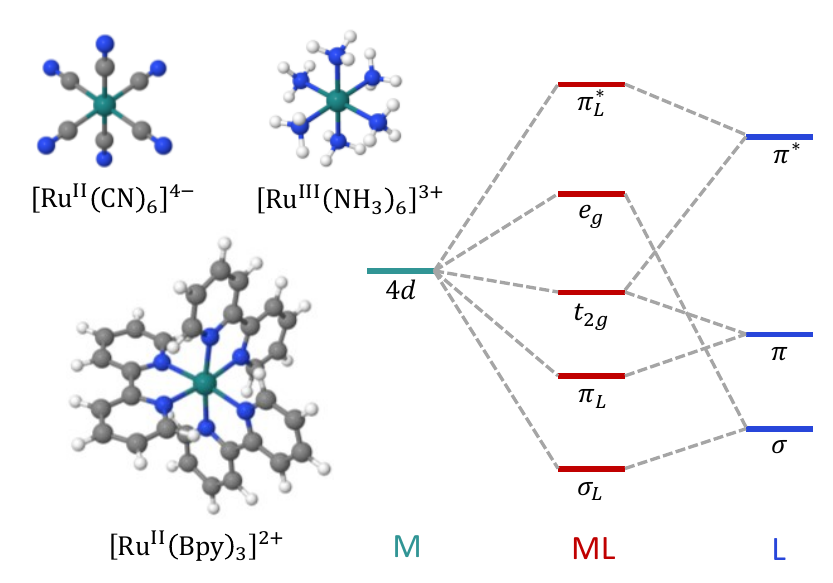}
\includegraphics[scale=1.0]{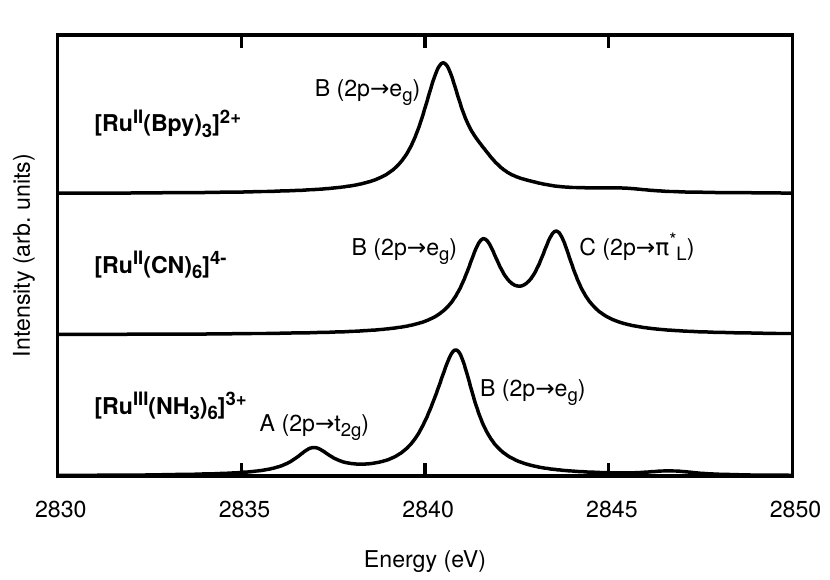} 
\label{FIG:absorption}
\end{figure}

In a 2p4d RIXS setup, the manifold of intermediate states is comprised of the L$_3$ excited states resulted by promoting an electron from the $2p_{3/2}$ orbitals in the metal center to valence states, whereas the manifold of final states are those in which a valence electron has been excited while leaving the metal $2p_{3/2}$ orbitals filled.
Thus, a natural starting point in understanding the 2p4d RIXS maps is to first analyze the metal L$_3$-edge absorption spectra, shown in Figure \ref{FIG:absorption}. 
Here, L$_3$-edge spectra were computed at the TDDFT/6-311G**+Sapporo-DKH-TZP-2012 level of theory employing the B3LYP exchange-correlation functional (see Computational Details). As it can be observed, the spectra for both \RuAm  and \RuCN  show two prominent features at 2837.0 and 2840.9 eV, and at 2841.6 and 2843.6 eV, respectively, while \RuBpy has a single dominant peak at 2840.6 eV. These features can be assigned to electronic dipole transitions from the Ru $2p_{3/2}$ orbitals into the unoccupied $t_{2g}$, $e_g$, and $\pi^*_L$ states as illustrated in Fig. \ref{FIG:absorption} by the simplified molecular orbital diagram for a generic octahedral system. The peaks associated with each of these unoccupied states are labelled as peaks A, B, and C, respectively. Note that since the Ru$^{\rm II}$ complexes are low-spin $d^6$ complexes, the $t_{2g}$ sates are completely filled, and thus, peak A is absent in their spectra. These results are in excellent agreement with the experimental results and previous calculations reported in Ref. \citenum{Van-Kuiken:2013:4444}, which validates our choice of exchange-correlation functional and basis sets.  

\begin{figure*}[!htpb]
\caption{Calculated Ru L$_3$-Valence RIXS maps at $\theta = 0^\circ$ for the \RuBpy, \RuCN, and \RuAm complexes. The colored dashed lines indicate the resonant absorption energies according to Fig. \ref{FIG:absorption}. A uniform shift of -2.3 eV has been applied in the absorption axis.}
\includegraphics[scale=1.0]{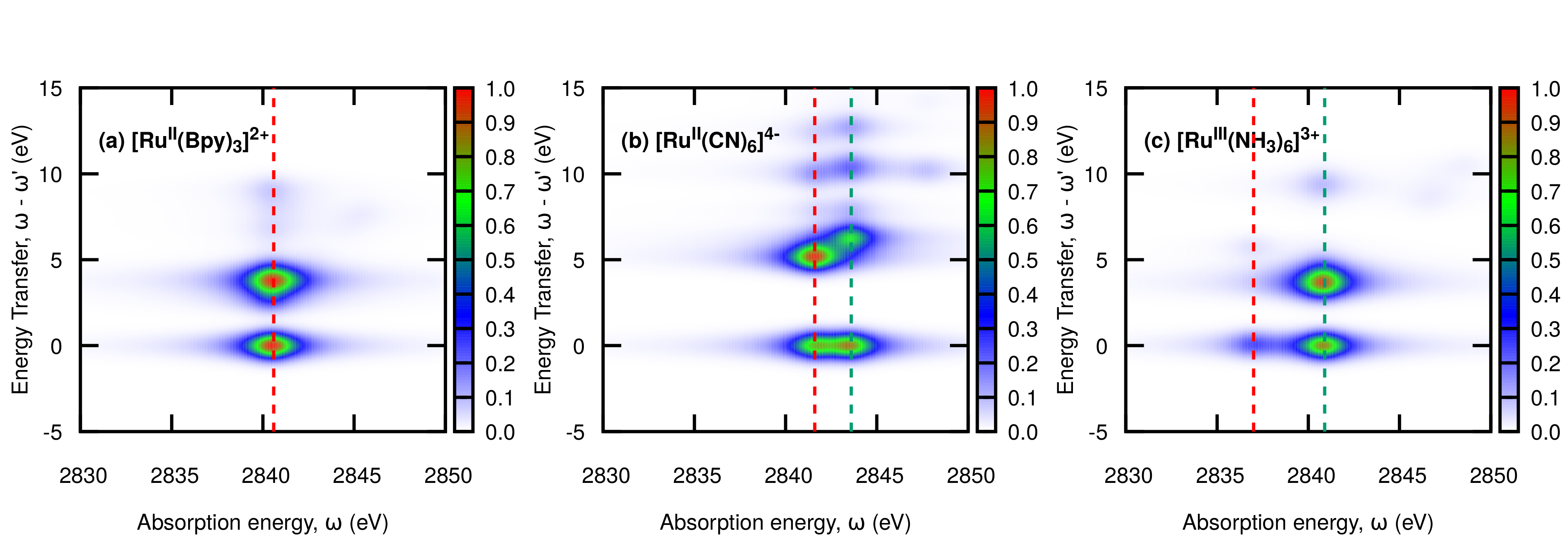} 
\label{FIG:maps}
\end{figure*}

\begin{figure*}[!htpb]
\caption{Experimental and calculated 2p4d RIXS slices along the resonant excitation energy axis for the \RuBpy, \RuCN, and \RuAm complexes.}
\includegraphics[scale=1.0]{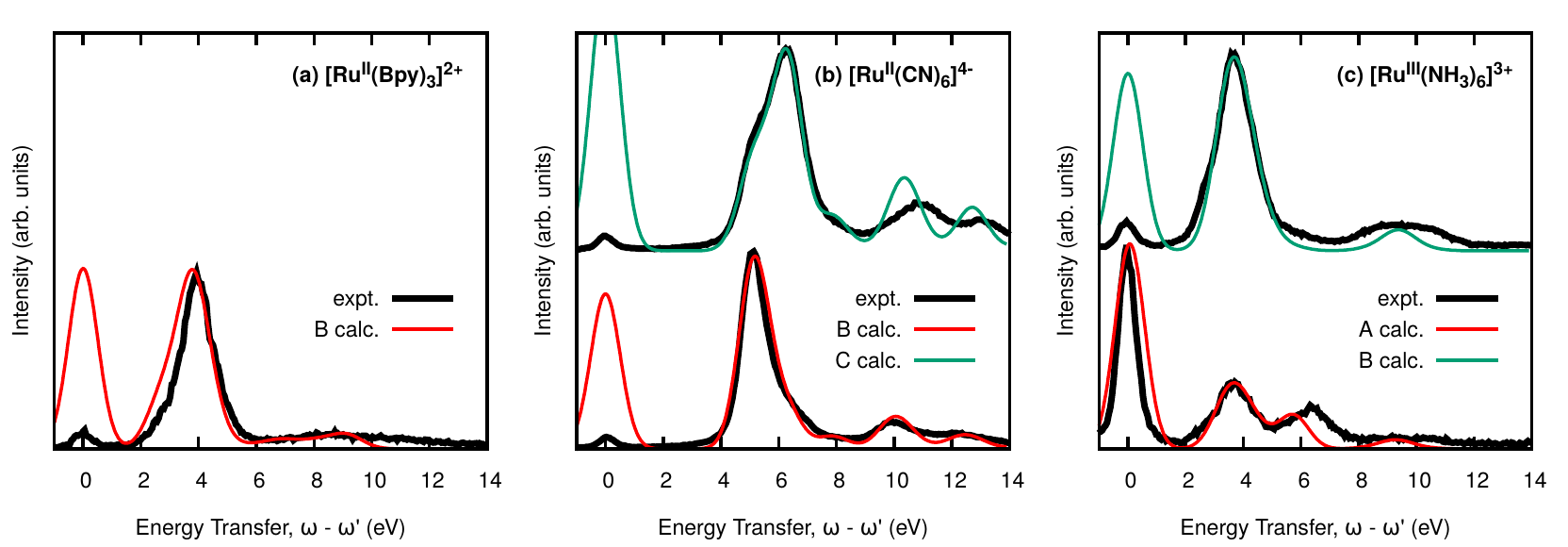} 
\label{FIG:slices}
\end{figure*}

In order to generate 2-D RIXS maps for these complexes, we computed the manifold of valence excited states and the corresponding 2p4d couplings (eq. \ref{EQN:excoupl}) at the same level of theory. The resulting maps (Figure \ref{FIG:maps}) are then constructed from eqs. \ref{EQN:KHdef} and \ref{EQN:rixs_ang} at $\theta=0^\circ$.
As suggested by their absorption spectra, the RIXS maps have strong features along the resonant L$_3$-edge absorption energies (marked by colored dashed lines in Fig. \ref{FIG:maps}). The features that appear with an energy transfer (defined as the difference between the absorbed and emitted energies) of 0 eV correspond to the elastic scattering process where the final state is the ground state itself. Here, we are interested in the inelastic features, where the final state is an excited valence state. These features appear at $\omega - \omega' > 0$. In order to facilitate analysis and comparison with experimental results, the slices along the colored dashed lines as well as the experimental traces are presented in Figure \ref{FIG:slices}.  
The first thing that can be observed in Fig. \ref{FIG:slices} is the excellent agreement between the experimental and calculated traces for all three complexes, with the clear exception of the elastic peaks, which are largely overestimated by our calculations. These discrepancies might be explained by several effects that were either not accounted for in the theoretical description, or difficult to control in the experiment. Such effects may include vibrational effects, Thomson scattering, surface scattering, and self-absorption effects. \cite{Kunnus:2016:7182}. Nonetheless, since our focus is on the description of the inelastic features, any further discussion about these effects will be deferred to a later work.

{\bf \RuBpy}: Both experimental and calculated traces (Fig. \ref{FIG:slices}a) show a single prominent inelastic peak at an energy transfer of 3.8 eV corresponding to the $(Ru2p)^5 (t_{2g})^6 (e_g)^1 \to (Ru2p)^6 (t_{2g})^5 (e_g)^1 $ emission and a substantially broad feature centered at 9.1 eV which can be attributed to the emission into the manifold of closely-spaced excited valence states with configuration $(Ru2p)^6 (\pi_L)^{n-1} (e_g)^1$. Note that in order to capture this broad feature in its entirety, an excessively large amount of roots must be computed. For this system, we considered 1300 valence excited states which is able to account for up to 10 eV in the energy transfer axis.

{\bf \RuCN}: The trace arising from the excitation at the B peak (Fig. \ref{FIG:slices}b red curve) shows 3 isolated features. The main feature at 5.2 eV can be assigned to the $(Ru2p)^5 (t_{2g})^6 (e_g)^1 \to (Ru2p)^6 (t_{2g})^5 (e_g)^1$ emission, whereas the features at 10.1 and 12.2 eV correspond to emissions into the $(Ru2p)^6 (\pi_L)^{n-1} (e_g)^1$ and $(Ru2p)^6 (\sigma_L)^{n-1} (e_g)^1$ excited valence states, respectively. 

The trace resulting from the excitation at the C peak (Fig. \ref{FIG:slices}b green curve) shows similar qualitative features with the main peak [$(Ru2p)^5 (t_{2g})^6 (\pi^*_L)^1 \to (Ru2p)^6 (t_{2g})^5 (\pi^*_L)^1$] at 6.2 eV followed by the emission features to the $(Ru2p)^6 (\pi_L)^{n-1} (\pi^*_L)^1$ and $(Ru2p)^6 (\sigma_L)^{n-1} (\pi^*_L)^1$ states at 10.4 and 12.7 eV, respectively. It is also possible to notice an intense shoulder at 5.2 eV (present in both the experiment and calculation), which originates primarily from the absorption at the B peak due to the significant overlap of both absorption features in \RuCN (see Fig. \ref{FIG:absorption} middle curve). Consequently, one can also observe the main feature in the C trace (at 7.8 eV) appearing as a shoulder in the B trace at the same energy.          

{\bf \RuAm}: The electronic structure of the \RuAm complex is significantly simpler than the other two complexes discussed earlier since the ammine  ligand has only $\sigma$-type bonds. The trace resulting from excitation at the A peak (Fig. \ref{FIG:slices}c red curve) shows two prominent features at 3.7 and 5.7 eV and a small peak at 9.1 eV. The latter two features can be attributed to the $(Ru2p)^5 (t_{2g})^6 \to (Ru2p)^6 (\sigma_L)^{n-1} (t_{2g})^6$ and $(Ru2p)^5 (t_{2g})^6 \to (Ru2p)^6 (N2p)^{n-1} (t_{2g})^6 $ emissions, respectively, while the feature at 3.7 eV is, in fact, a shoulder resulting from excitation at the B peak. Accordingly, the main feature for the B trace corresponding to the $(Ru2p)^5 (t_{2g})^5 (e_g)^1 \to (Ru2p)^6 (t_{2g})^4 (e_g)^1$ emission can be observed at 3.7 eV followed by a small peak at 9.4 eV assigned to the emission into the $(Ru2p)^6 (\sigma_L)^{n-1} (e_g)^1$ states.

In summary, we have proposed a  pseudo-wavefunction-based protocol to compute couplings between two different manifolds of excited states relative to a common ground state within the context of TDDFT. These couplings are the required matrix elements to solve the Kramers--Heisenberg formula for RIXS and several other types of two-photon spectroscopies \cite{Gelmukhanov:1999:87,deGroot:2001:1779,Norman:2018:7208}. We demonstrated the validity of the approach by calculating the 2p4d RIXS maps of three prototypical ruthenium complexes and comparing them to experimental spectra. The overall qualitative features are well reproduced by the computations and the relative transfer energies for all three systems are correct to within 0.6 eV. Thus, the present method has the potential to become a useful tool for the prediction and interpretation of two-photon experiments currently being undertaken at X-ray light source facilities. 

\section{Computational Details}

The ground-state geometries were previously optimized using the ORCA quantum chemistry package \cite{Neese:2018:e1327} at the B3LYP/def2-TZVP level of theory. Solvent effects were modeled via the conductor-like polarizable continuum model (CPCM) \cite{Barone:1998:1995}.

Excited-state computations were performed within the Tamm--Dancoff approximation \cite{Hirata:1999:291} using a development version of the NWChem package \cite{Valiev:2010:1477,Apra:2012:184102}. All computations employed the B3LYP functional \cite{Becke:1988:3098,Lee:1988:785}, the Sapporo-DKH3-TZP-2012 basis set \cite{Noro:2012:1124} for the Ru atoms and the 6-311G** basis for all the remaining atoms. Solvent (water) effects were included implicitly via the Conductor-like Screening Model (COSMO) \cite{Klamt:1993:799,York:1999:11060}, and scalar relativistic effects were included via the Zeroth-order Regular Approximation (ZORA) model potential of van Lenthe et al. \cite{van-Lenthe:1994:9783,Van-Wullen:1998:392,Nichols:2009:491}. Spin-orbit splittings have been neglected. We have previously shown that spin-orbit coupling does not strongly influence the L$_3$-edges of Ru complexes.\cite{Van-Kuiken:2013:4444} In the atomic $J$-$J$ coupling picture, transitions follow the $\Delta J = 0, \pm 1$ dipole selection rule. This suggests that excitation (and de-excitation) transitions involving the L$_3$-edge are allowed to (and from) all d orbitals. These transitions are dipole allowed in the spin-free picture, yielding the same selection rule. 

In order to generate the manifold of Ru L$_3$-edge excited states, the energy window from -105.0 to -103.0 $E_h$ was chosen and a total of 200 roots were computed for each complex, while the manifold of valence excited states employed no window restriction and was comprised of 200 roots for the \RuAm complex and 1300 roots for \RuBpy and \RuCN. 

In implementing eq. \ref{EQN:KHdef}, a uniform lifetime broadening, $\Gamma_n$, of 2.4 eV was assumed, and the Dirac delta was approximated by a Gaussian function with a FWHM of 1.2 eV to account for the combined effects of experimental resolution, final lifetime state broadening, and vibronic effects. 
 
\vspace{0.5cm}
{\bf Acknowledgments:}
\input{acknowledgment}

\bibliography{main}

\end{document}

%% file: acknowledgment.tex
This work was supported by the U.S. Department of Energy, Office of Science, Office of Basic Energy Sciences, Chemical Sciences, Geosciences and Biosciences Division under Award Nos. KC-030105172685 (D.R.N., N.G.), DE-AC02-76SF00515 (E.B.), DE-SC0019277 (B.P. and M.K.).  B.P. acknowledges support by the NSF GRFP (No. DGE-1762114). This research benefited from computational resources provided by EMSL, a DOE Office of Science User Facility sponsored by the Office of Biological and Environmental Research and located at PNNL. PNNL is operated by Battelle Memorial Institute for the United States Department of Energy under DOE Contract No. DE-AC05-76RL1830. This research also used resources of the National Energy Research Scientific Computing Center (NERSC), a U.S. Department of Energy Office of Science User Facility operated under Contract No. DE-AC02-05CH11231. Use of the Stanford Synchrotron Radiation Lightsource, SLAC National Accelerator Laboratory, is supported by the U.S. Department of Energy, Office of Science, Office of Basic Energy Sciences under Contract No. DE-AC02-76SF00515.